\documentclass[preprintnumbers,twocolumn,floats,aps,nofootinbib]{revtex4-1}
\pdfoutput=1
\usepackage{graphicx}
\usepackage{color}
\usepackage{amsmath,amssymb,amsfonts}
\usepackage{verbatim}   
\usepackage{subfigure}  
\usepackage{acronym}
\usepackage{hyperref}
\usepackage{mathrsfs}
\usepackage{multirow}
 \usepackage{cancel}
 \usepackage{url}
  \usepackage{slashed}
 \usepackage{hyperref}

\def\beq{\begin{equation}\begin{aligned}}
\def\eeq{\end{aligned}\end{equation}}

\def\OO{\mathcal{O}}

\begin{document}

\title{X-ray lines from $R$-parity violating decays of keV sparticles}

\author{Christopher Kolda}
\email{ckolda@nd.edu}
\author{James Unwin}
\email{junwin@nd.edu}
\affiliation{Department of Physics, University of Notre Dame, Notre Dame,
IN, 46556, USA}

\begin{abstract}

If $R$-parity is only mildly violated then the lightest supersymmetric particle (LSP) can be stable over cosmologically time-scales and still account for the dark matter relic density. We examine the possibility of generating detectable X-ray lines from $R$-parity violating decays of keV-scale LSP dark matter to neutrino-photon pairs.  Specifically, we consider scenarios in which the LSP is a light gravitino, bino, or hidden sector photino.  Potential signals are discussed in the context of recent claims of an unidentified 3.5 keV X-ray line in studies of stacked galaxy clusters. We comment on the difficulties in obtaining the observed relic density for keV scale bino or hidden photino dark matter and some possible resolutions.

\end{abstract}

\date{March 25, 2014}
\maketitle


\section{Introduction}

Despite negative searches at the LHC thus far, TeV scale supersymmetry (SUSY) remains the leading resolution to the hierarchy problem, particularly in light of the discovery of a Higgs boson not much above the $Z$-mass. 
Whilst experimental searches and theoretical considerations suggest that most of the SUSY spectrum should be above the weak scale, it is quite conceivable that certain neutral states could be substantially lighter. Such light sparticles could have interesting cosmological and astrophysical implications. Here we shall focus on the prospect of keV-scale SUSY states which can decay in such a manner as to produce observable X-ray signals. 
We are inspired to think about decays of light particles by the tentative 3.5 keV line observed in the combined spectra of multiple galaxy clusters as studied by the XMM-Newton X-ray observatory  \cite{Bulbul:2014sua,Boyarsky:2014jta}. Thus we shall typically phrase our discussion in terms of this benchmark point and examine models of dark matter which might accommodate this phenomenon. This signal can be interpreted in terms dark matter which decays to a photon and an (effectively) massless degree of freedom with a mass and lifetime around
\beq
m_{\rm DM} & \simeq7~{\rm keV}~,\\
\tau_{\rm DM} & \simeq2\times10^{27}~\text{-}~2\times10^{28}~{\rm s}~.
\label{1}
\eeq
Some potential models have been proposed which might account for this observational anomaly \cite{Bulbul:2014sua,Ishida:2014dlp,Finkbeiner:2014sja,axion,Jaeckel:2014qea,  Abazajian:2014gza,Krall:2014dba,Lee:2014xua,   Baek:2014qwa,Choi:2014tva,Nakayama:2014ova, Frandsen:2014lfa,Aisati:2014nda,Kong:2014gea,Cicoli}, the possibility of decaying sterile neutrinos or axions receiving particular attention. Given the wide expectation that SUSY should play a leading role in physics beyond the Standard Model, it is interesting to explore possible SUSY explanations for this  signal.

As is well known, the canonical SUSY extension of the Standard Model, the MSSM, requires the imposition of an ({\em ad hoc}) discrete $Z_2$ symmetry: $R$-parity
\beq
(-1)^{3(B-L)+2s}~.
\eeq
 Under this discrete symmetry the Standard Model (superpartner) states transform as even (odd) representations. This is necessary purely for phenomenological purposes since  there are  dimension four and five operators of the form 
\beq
\mu' \boldsymbol {L H_u}~,\quad
\lambda  \boldsymbol{L L \overline{E}}~,\quad
\lambda' \boldsymbol{L Q \overline{D}}~,\quad
\lambda'' {\boldsymbol {\overline{U}\overline{D}\overline{D}}}~,
\label{rpv}
\eeq
which, unless the couplings are small, are problematic as they lead to fast proton decay in conflict with experimental searches  \cite{Abe:2011ts,Hikasa:1992je}. 
If $R$-parity is an exact symmetry then the lightest supersymmetric particle (LSP) is stable. 
Intriguingly a stable LSP can play the role of dark matter and the occurrence of a well motivated dark matter candidate is arguably one of the great triumphs of SUSY extensions of the Standard Model, see e.g.~\cite{Jungman:1995df}.

An interesting variation is the scenario in which $R$-parity is mildly violated  \cite{Hall:1983id,Dawson:1985vr,Barbier:2004ez,Dreiner:1997uz} such that the LSP is effectively  stable on cosmologically time-scales and can still account for the dark matter. However, a small fraction of these states will decay presently and for an appropriate lifetime can potentially generate detectable signals. 
Good candidates for the LSP in the MSSM are the fermion superpartners to the known boson fields, such as the neutralino or gravitino. Assuming mild $R$-parity violation (RPV), a fermion LSP lighter than the electron will dominantly decay to a photon-neutrino pair, unless the spectrum is supplemented with additional light fermion states.
For an LSP decaying to a photon and an effectively massless state to produce X-ray signals, the parent state must be around the  keV scale. More specifically,  to match the recent anomaly at 3.5 keV \cite{Bulbul:2014sua} the parent state should be roughly 7 keV. Since we suppose that the LSP constitutes the dark matter, this will be `warm' dark matter. It should be noted that sub-keV thermally produced `hot' dark matter leads to the erasure of density perturbations at scales shorter than its free streaming length and is in conflict with observations of small scale structure, see e.g.~\cite{Viel:2005qj}.

Although the cluster anomaly at 3.5 keV might be regarded as tentative at this stage, it provides motivation for us to explore interesting, non-standard, scenarios of SUSY. We begin  in Sect.~\ref{S2}  by investigating if a keV gravitino can give rise to decay signals and show that generically this is not the case. In Sect.~\ref{S3}, we explore the prospect of generating X-ray lines from decaying bino dark matter and argue that this is quite possible. However, as we discuss, obtaining the correct relic density for bino dark matter requires significant model building. In Sect.~\ref{S4}, we examine the motivation for light hidden sector photini and argue that such a state could be the LSP. We show that a 7 keV hidden photino LSP can have a suitable abundance to match the observed dark matter relic density and give rise to the 3.5 keV cluster line via $R$-parity violating decays.


\section{Decaying Gravitino LSP}
\label{S2}

The archetypal example of a motivated light superpartner which can, in principle, decay on cosmological timescales is the gravitino. 
The gravitino is an integral aspect of supersymmetric extensions of the Standard Model. The mass of this state, which is tied to the scale of $F$-term SUSY breaking, is an unknown parameter and could vary over a large range.  Once SUSY is broken the effects of this breaking will generically be mediated via Planck suppressed operators. However, if the superpartners of the Standard Model states feel additional sources of SUSY breaking, then the gravitino can be significantly lighter than the rest of the superpartner spectrum. The typical scale of the SUSY soft masses due to gauge mediation is
\beq
\widetilde{m}\sim\frac{F}{M}~,
\label{GM}
\eeq
where $M$ is the mass scale of the SUSY breaking mediators and $F$ is a SUSY breaking F-term. Whereas the gravitino mass is set by gravity mediation
\beq
m_{3/2}=\frac{F}{\sqrt{3}M_{\rm Pl}}~.
\label{3/2}
\eeq
This permits for large separations in the mass scales $m_{3/2}\ll m_{\widetilde{f}}$ provided $M\ll M_{\rm Pl}$.
Whilst the gravitino is a well motivated candidate, we find that its lifetime is typically too long to reproduce the anomaly of interest.

A gravitino with mass $m_{3/2} < m_e$ in the MSSM will decay via ${\widetilde G}\rightarrow\nu\gamma$ in the presence of $R$-parity violation. The details of the LSP decay depend on how the dominant source $R$-parity is introduced and decays due to the presence of either the bilinear or trilinear RPV operator of eq.~(\ref{rpv}) with sizeable couplings are of particular interest, see e.g.~\cite{Takayama:2000uz,Ishiwata:2008cu,Bobrovskyi:2010ps,Endo:2009by,Buchmuller:2007ui,Bajc:2010qj}.

Parameterising the details of the RPV process which results in the gravitino decay in terms of a suppression factor $\mathcal{F}$, the lifetime can be generally expressed as follows \cite{Takayama:2000uz,Ishiwata:2008cu}
\beq
\tau_{\widetilde{G}}=\left(
\frac{\mathcal{F}}{16\pi}\frac{m_{3/2}^3}{M_{\rm Pl}^2}\right)^{-1}
\simeq6\times10^{29}~{\rm s}\left(\frac{1}{\mathcal{F}}\right)
\left(\frac{7~{\rm keV}}{m_{3/2}}\right)^3~.
\eeq
Typically $\mathcal{F}\ll1$; for instance, in the case of $R$-parity violation due the bilinear operator $\boldsymbol{LH_u}$, the gravitino can decay to $\nu\gamma$ via a neutrino-bino mixing term and this factor is parametrically $\mathcal{F}\sim (m_\nu/m_{\widetilde{B}})\ll1$. 
However, even in the extreme limit $\mathcal{F}\simeq1$ the lifetime is still too long to account for the claimed 3.5 keV line of \cite{Bulbul:2014sua}.\footnote{It was subsequently suggested \cite{Bomark:2014yja} that gravitinos decaying radiatively via a trilinear RPV coupling could account for the signal. However, this result is based on an erroneous expression for the decay rate in \cite{Lola:2007rw}. For the correct form see e.g.~\cite{Bajc:2010qj}, which conforms with our general conclusion.} It is interesting to note that for $m_{3/2}\lesssim$ few keV the gravitino is essentially stable for all phenomenological purposes.


\begin{figure*}[t!]
\begin{center}
\includegraphics[height=40mm]{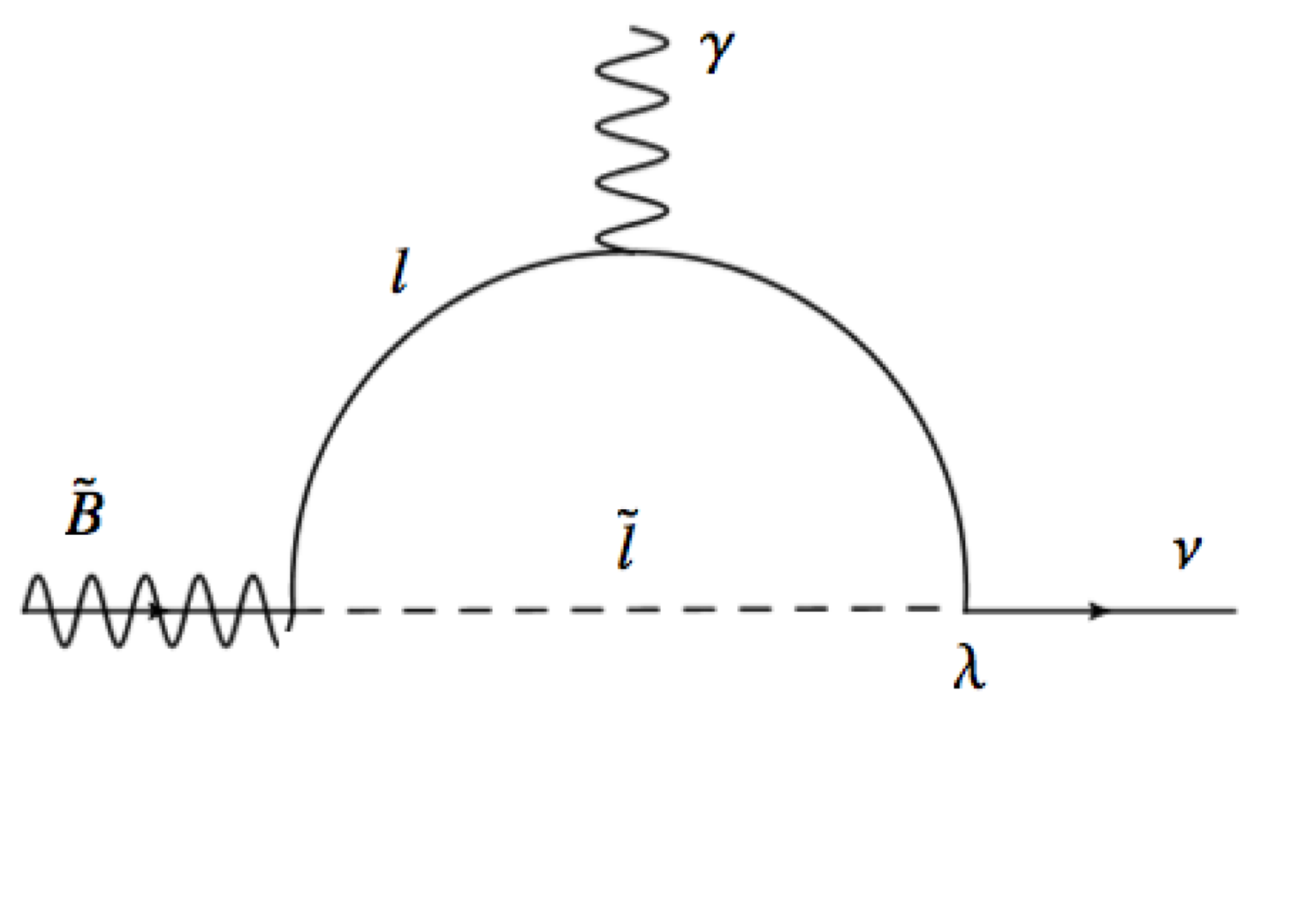}
\hspace{1mm}
\includegraphics[height=42mm]{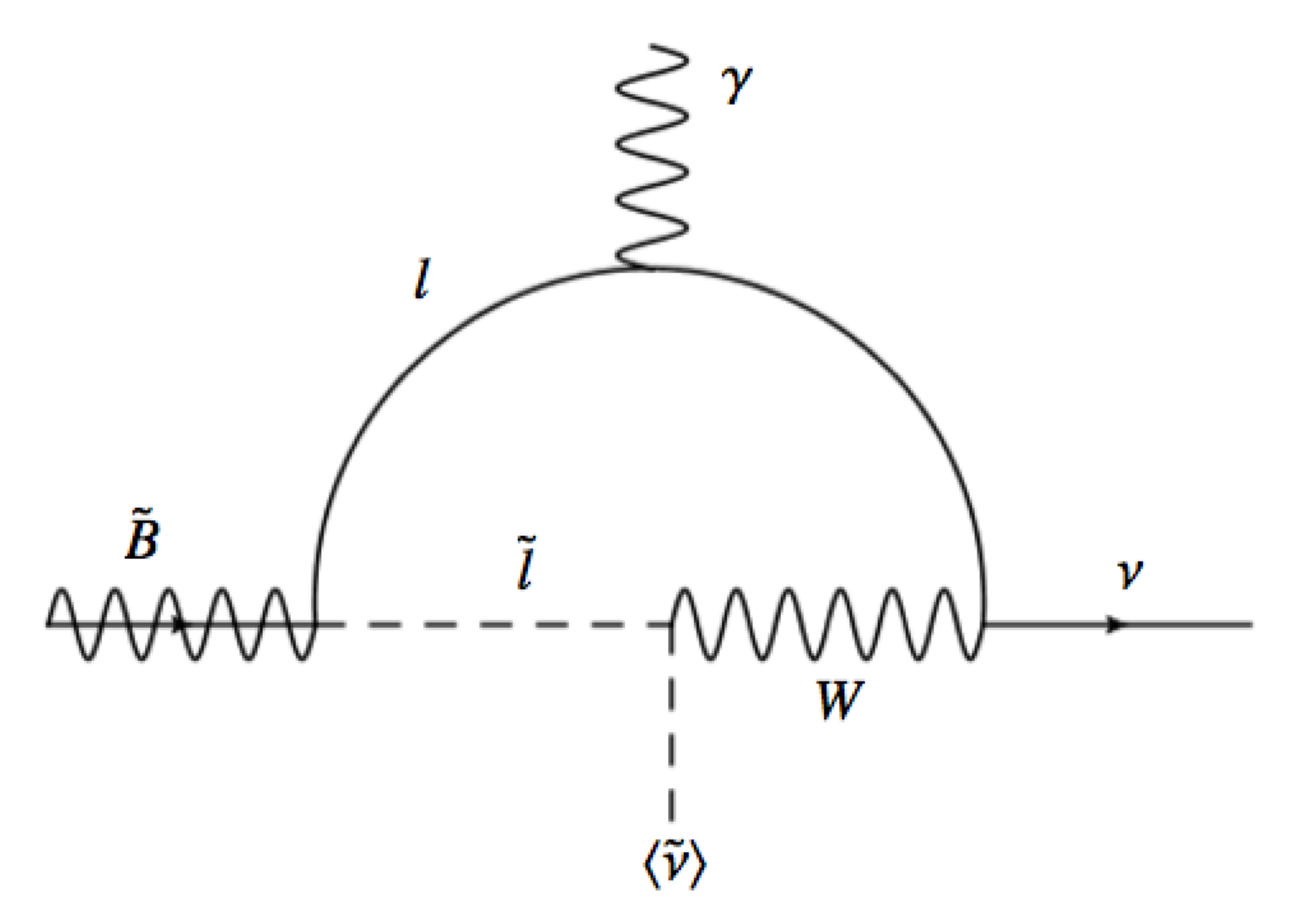}
\hspace{1mm}
\includegraphics[height=38mm]{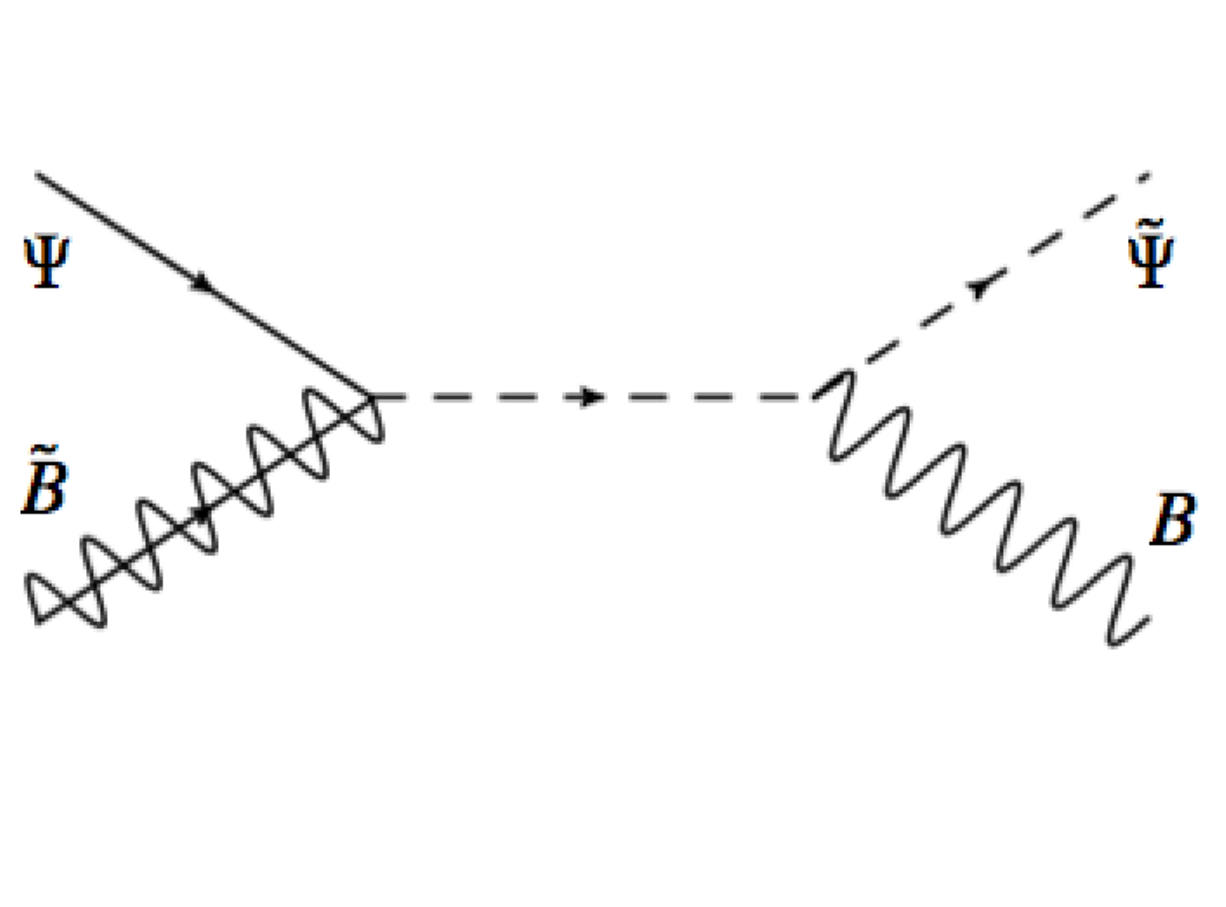}
\caption{Bino LSP decay involving the trilinear $R$-parity violating operator $\lambda \boldsymbol{LLE}$ (left) and due to a sneutrino VEV (centre). The bino assimilation mechanism of \cite{D'Eramo:2011ec} (right).
 \label{fig}}
\end{center}
\end{figure*}

\section{Decaying Light Bino LSP}
\label{S3}

Given that the gravitino scenario is seemingly unviable, we shall consider alternative LSP candidates which could generate this signal through decay processes. Specifically, we shall focus on models in which the LSP is a light bino or hidden sector photino.

It is traditional to constrain the values of the soft masses with GUT boundary conditions, which fix the ratios of the SUSY breaking gaugino Majorana masses
\beq
M_1=\frac{5}{3}\tan^2\theta_W M_2\simeq\frac{1}{2}{M_2}~.
\label{M1}
\eeq
This typically implies that the binos and winos should receive comparable masses, and experimental bounds under this assumption limit the mass of the lightest neutralino to be in excess of around 50 GeV \cite{Hikasa:1992je}. If the condition eq.~(\ref{M1}) is not imposed then there is a great deal more freedom in the relative masses of the electroweakino states. In particular, it is possible to tune the soft masses $\mu$, $M_1$ and $M_2$ such that the neutralino mass matrix develops a vanishing eigenvalue and thus the lightest neutralino can potentially be ultra-light or even massless \cite{Profumo:2008yg,Gogoladze:2002xp,Dreiner:2009ic,Davies:2011mp}. For the MSSM a vanishing eigenvalue for the neutralino mass matrix occurs for 
\beq
M_1=\frac{M_2M_Z^2\sin(2\beta)\sin^2\theta_W}{\mu M_2-m_Z^2\sin(2\beta)\cos^2\theta_W}~.
\eeq
Notably, if this light neutralino is essentially pure bino, as is typical, it circumvents the current direct search bounds over much of parameter space \cite{Dreiner:2009ic}.

A light bino LSP can be realised in an alternative manner, which does not require tuning of the MSSM electroweak soft masses. Gaugino Majorana masses can be forbidden at leading order by an $R$-symmetry  broken only by $M_{\rm Pl}$-scale operators, in which case the gauginos will typically be very light. However, if the MSSM spectrum is supplemented with additional chiral superfields transforming as an SU(2) triplet $\boldsymbol T$ and an SU(3) octet $\boldsymbol O$ then the winos and gluinos can acquire masses via Dirac terms of the form \cite{Fox:2002bu}
\beq
\int{\rm d}^2\theta\frac{\boldsymbol W_\alpha'}{M}\left[\lambda_G{\rm Tr}\left(\boldsymbol{OG}^\alpha\right)+\lambda_W{\rm Tr}\left(\boldsymbol{TW}^\alpha\right)\right]~,
\eeq
where $\boldsymbol{W'}$ is a spurion field which develops a $D$-term. Thus, the winos and gluino receive masses of order $m_{2,3}=\lambda_{W,G} D'/M$ and for appropriate sizes of  the D-term breaking and mediation scale these can be phenomenologically viable (i.e.~TeV scale). In the absence of a gauge singlet chiral superfield field (which are often deemed undesirable, as they can result in tadpole problems, see e.g.~\cite{Ellwanger:2009dp}) the bino will only receive a mass contribution from anomaly mediation \cite{Giudice:1998xp,Randall:1998uk}
 \beq
m_{\widetilde{B}}\sim\frac{g^2b_1}{16\pi^2}m_{3/2}\sim 7~{\rm keV}\left(\frac{m_{3/2}}{1~{\rm MeV}}\right)~.
\label{AM}
 \eeq
 In this manner a natural hierarchy emerges between the bino and the other sparticles, as in \cite{Davies:2011mp}, due only to the spectrum and symmetry structure.
It should be noted that an MeV gravitino is cosmologically stable and, hence, also gives a contribution to the dark matter relic density of order \cite{de Gouvea:1997tn}
\beq
\Omega_{3/2}h^2\sim0.01\left(\frac{1~{\rm MeV}}{m_{3/2}}\right)
\left(\frac{T_{\rm RH}}{1~{\rm TeV}}\right)
\left(\frac{\widetilde{m}}{1~{\rm TeV}}\right)^2~.
\eeq
If the dark matter relic density has significant contributions from both bino and gravitino components, then the lifetime of the decaying bino LSP must be scaled linearly, in terms of its fractional contribution to the relic density, in order match the observed flux. For simplicity we shall focus on scenarios in which the bino accounts for essentially all of the dark matter relic density. This case arises, for instance, if the reheat temperature is low $T_{\rm RH}\ll 1$ TeV, such that only a small abundance of gravitinos is thermally produced.

A keV scale bino LSP can decay at loop level in the presence of one of the $R$-parity  violating trilinear operators of eq.~(\ref{rpv}). Specifically, we consider the case involving the leptonic operator  $\lambda \boldsymbol{LLE}$, leading to the diagram illustrated in Fig.~\ref{fig}: left panel. The lifetime of the bino decaying $\widetilde{B}\rightarrow\nu\gamma$ is given by \cite{Dreiner:2009ic,Dawson:1985vr}  
\beq
\tau_{\widetilde{B}}\simeq5\times10^{27}~{\rm s}
\left(\frac{10^{-8}}{\lambda}\right)^2
\left(\frac{m_{\widetilde{f}}}{2~{\rm TeV}}\right)^4
\left(\frac{7~{\rm keV}}{m_{\widetilde{B}}}\right)^3.
\label{11}
\eeq
Note that experimental limits bound the coupling to be $\lambda\lesssim0.05$, assuming that this is the sole source of $R$-parity  violation  \cite{Barbier:2004ez,Dreiner:1997uz}. In the presence of multiple operators these limits can become substantially stronger \cite{Allanach:1999ic}. Thus provided that $R$-parity is only violated slightly then a 7 keV bino can be cosmologically stable and present a decay rate suitable to explain the 3.5 keV cluster line. Settings in which certain $R$-parity violating operators occur with small coefficients have been suggested in the literature, e.g.~\cite{Buchmuller:2007ui,Endo:2009by,Bobrovskyi:2010ps,Barger:2008wn}.

The above trilinear RPV operator will typically induce a sneutrino VEV  $\langle\widetilde{\nu}\rangle\neq0$ which can provide an alternative decay process, as in Fig.\ref{fig}: centre panel. In eq.~(\ref{11}) we have assumed that this VEV is small such that the RPV operator $\boldsymbol{LLE}$ sets the bino lifetime. Further, sneutrino VEVs can arise in alternative manners, for instance via the RPV bilinear $\mu'\boldsymbol{LH_u}$. The magnitude of the VEV is model dependent and linked to the associated RPV coupling constant which can (and is often required to) be small, see e.g.~\cite{Barbier:2004ez}.  For the case of the bilinear operator $\mu'\boldsymbol{LH_u}$ a sneutrino VEV is generated of order \cite{Hall:1983id}
\begin{align}
\langle \widetilde{\nu}\rangle 
&\simeq \left(\frac{\mu'}{m_{3/2}}\right)^2v\cos\beta\\
&\simeq
 10~{\rm eV}\times\left(\frac{\mu'}{10~{\rm eV}}\right)^2 \left(\frac{1~{\rm MeV}}{m_{3/2}}\right)^2\left(\frac{\cos\beta}{0.5}\right)~.
\notag
\end{align}

Accordingly, the lifetime of a bino decaying via the diagram of Fig.~\ref{fig}: centre panel is given by \cite{Dawson:1985vr} 
\beq
\tau'_{\widetilde{B}}
\simeq 
10^{28}~{\rm s}
\left(\frac{m_{\widetilde{l}}}{1~{\rm TeV}}\right)^4
\left(\frac{10~{\rm eV}}{\langle\widetilde{\nu}\rangle}\right)^2
\left(\frac{10~{\rm keV}}{m_{\widetilde{B}}}\right)^3.
\label{snu}
\eeq
Thus a lifetime  appropriate  to match the aforementioned cluster anomaly can be found for reasonable values of the slepton masses and sneutrino VEV, with some parameter freedom. Note that the sneutrino VEV generates a mass contribution for the neutrino (assuming Majorana gauginos) of order \cite{Hall:1983id} (see also \cite{Takayama:1999pc})
\beq
\Delta m_\nu
&\simeq g^2 \langle \widetilde \nu \rangle^2/m_{\widetilde B}\\
&\simeq10^{-3}~{\rm eV}
\left(\frac{\langle \widetilde \nu \rangle}{10~\rm{eV}}\right)^2
\left(\frac{10~{\rm keV}}{m_{\widetilde B}}\right)~.
\label{nu}
\eeq
Observational limits on the neutrino masses ($m_{\nu}\lesssim1$ eV) constrain the size of the sneutrino VEV.

One potential difficultly is that, as the bino annihilation rate is low, if it is produced thermally then it is typically overproduced and consequently it is challenging to realise the observed relic density via freeze out. Furthermore, it is unlikely that the relic density can be reduced via tuning the reheat temperature \cite{Giudice:2000ex} to below the mass of the bino $T_{\rm RH}\lesssim {\rm keV}$, as observations indicate that the temperature of the universe was in excess of a few MeV. This is supported by the successful predictions of the primordial abundances of nuclei in theories of big bang nucleosynthesis, see e.g~\cite{Jedamzik:2006xz}.
One manner to potentially achieve the correct density of bino dark matter is the {\em assimilation} mechanism of D'Eramo, Fei and Thaler  \cite{D'Eramo:2011ec}. In this scenario the MSSM is supplemented with new exotic states $\Psi$ which carry a particle asymmetry such that their number density remains considerable at temperatures below their mass. These exotic states co-annihilate efficiently with the bino LSP via $\widetilde{B}\Psi\rightarrow\widetilde{\Psi}B$ (assimilation, see Fig.~\ref{fig}: right panel) and $\widetilde{B}\widetilde{\Psi}\rightarrow\Psi B$ (destruction), where $B$ is the hypercharge gauge boson. Subsequently, late decays of $\widetilde{\Psi}$ to the bino LSP, which may violate the global quantum number carried by the $\Psi$, are responsible for setting the bino abundance. Clearly the $\Psi$ must be sufficiently long lived such that they do not decay before the bino abundance is suitably depleted. Since the production of binos from the thermal bath is only suppressed once the bath cools to below the bino mass, the $\Psi$ must survive to below $T\lesssim m_{\widetilde{B}}/25$. For a 7 keV bino this corresponds to a temperature around 300 eV. If these states decay after recombination ($T\lesssim$ 1 eV) then this can lead to CMB signals which are strongly constrained, see e.g.~\cite{Kamionkowski:1999qc}. Further, there are stringent bounds on hadronic decays of the $\Psi$ or $\widetilde{\Psi}$ from measurements of big bang nucleosynthesis observables \cite{Jedamzik:2006xz}. For non-hadronic decays of the  $\Psi$ and $\widetilde{\Psi}$ prior to recombination, constraints on energy injection can be satisfied, in particular for the case $\widetilde{\Psi}\rightarrow\nu\widetilde{B}$. Thus viable models can be constructed with bino abundances appropriate to account for the dark matter relic density. Also there could be a potentially observable deviation in the number of additional relativistic species $N_{\rm eff}$ due to the binos remaining in thermal equilibrium later than neutrino decoupling, see e.g.~\cite{Boehm:2013jpa}.

Under the assumption that the $\widetilde{\Psi}$ is the NLSP and each $\widetilde{\Psi}$ decay produces a single bino LSP, an estimate for the required particle asymmetry $\eta_{\widetilde{\Psi}}\equiv n_{\widetilde{\Psi}}-n_{\overline{\widetilde{\Psi}}}$ in order to generate the dark matter relic density can be found from the relation between the  $\widetilde{\Psi}$ asymmetry and the bino mass
\beq
\frac{\Omega_{\rm DM}}{\Omega_{\rm B}}\simeq \frac{m_{\widetilde{B}}~\eta_{\widetilde{\Psi}}}{m_p~\eta_B}~,
\eeq
where $m_p\simeq1$ GeV is the proton mass and $\eta_B\simeq6\times10^{-10}$ is the baryon asymmetry. The ratio of relic densities is determined to be $\Omega_{\rm DM}/\Omega_{\rm B}\simeq5$, hence for $m_{\widetilde{B}}\simeq7$ keV this implies
\beq
\eta_{\widetilde{\Psi}}\simeq~10^{6}\times\eta_B\simeq 6\times10^{-4}~.
\eeq
Thus this scenario requires a significantly larger asymmetry than that associated to baryon number. This is a mild departure to the original model of  \cite{D'Eramo:2011ec} in which it was envisaged that baryon asymmetry would be generated via the late decays of the $\Psi$ and $\widetilde{\Psi}$ states, which is a requirement we do not retain here.
This setting provides one method for obtaining the correct relic abundance of bino dark matter and it is conceivable that alternative mechanisms might be constructed.

\section{Decaying Hidden Photino LSP}
\label{S4}

Given the moderate model building necessary to obtain the observed relic density for keV bino dark matter, we turn to the possibility that the LSP is not a standard MSSM state, but rather the superpartner of some U(1) vector boson sequestered from the visible sector. 
Indeed, the existence of (many) light hidden sector abelian vector bosons, and their associated superpartners, are motivated by string theory \cite{Arvanitaki:2009hb,Ibarra:2008kn}. In type IIB string theory each 3-cycle $\Sigma_i^3$, labelled by $i$, can be associated to a 4d vector field in terms of the Ramond-Ramond form $C_4$
\beq
A_\mu^i=\int_{\Sigma_i^3}C_4~.
\eeq
These vector bosons inherit a gauge symmetry from the 10d gauge symmetry of $C_4$. Typically in realistic string compactifications there are $\OO(10)$ 3-cycles, and as many associated gauge bosons. Whilst a large number of these gauge symmetries are likely broken in the UV, by fluxes or otherwise, some may remain unbroken in the low energy theory \cite{Arvanitaki:2009hb}. 
There can also be further hidden U(1) gauge symmetries arising from branes separated from the Standard Model brane stack on the compactification manifold. If these branes have only vector-like matter content with high scale masses (which is a common scenario), then the low energy theory contains an unbroken U(1) with no light charged states  \cite{Ibarra:2008kn}. 

After SUSY is broken and the effects of this breaking are mediated to the hidden sectors, the fermion partners to these hidden sector gauge bosons receive soft masses accordingly. 
Furthermore, it is natural for these hidden sector photini to be significantly lighter than visible sector sparticles if they only feel SUSY breaking via gravitational effects. In this case the photini will acquire masses near the gravitino mass $m_{3/2}$ of the form given in eq.~(\ref{3/2}) or, alternatively, eq.~(\ref{AM}). In contrast, the visible sector superpartners, including the gauginos, can obtain their masses via some other mediation mechanism leading to TeV soft masses in the visible sector, for example via gauge mediation as given in eq.~(\ref{GM}).

Although the hidden sectors may not communicate directly with the visible sector, as is well known, U(1) vector bosons will generically mix \cite{Holdom:1985ag}. For the case of two U(1)'s, this mixing appears in the gauge sector Lagrangian as follows \cite{Dienes:1996zr}
\beq
\mathcal{L}_{\rm gauge}= \frac{1}{32}\int{\rm d}^2\theta(W_aW_a+W_bW_b-2\chi W_aW_b)~,
\eeq
where $W_{a,b}$ are chiral gauge field strength superfields for each U(1) symmetry, defined by $W=\overline{D}^2DV$ in terms of $V$ the vector superfield. In supergravity such kinetic mixing between U(1)'s generically arises due to Planck suppressed operators and, from a stringy perspective, this occurs due to open strings stretching between the Standard Model brane stack and hidden sector $D$-branes \cite{Arvanitaki:2009hb,Ibarra:2008kn}. Thus it is expected that the abelian gauge symmetries, specifically U(1) hypercharge, will mix with any hidden sector U(1)'s. 

The pure gauge part of the Lagrangian can be rewritten in canonical form by making the following field redefinition \cite{Dienes:1996zr}
\beq
V_a^\mu\rightarrow V_a'{}^\mu=V_a^\mu-\chi V_b^\mu~,
\eeq
such that $W_a\rightarrow W'_a=W_a-\chi W_b$,
which diagonalises the gauge sector Lagrangian contribution
\beq
\mathcal{L}_{\rm gauge}= \frac{1}{32}\int{\rm d}^2\theta(W_a'W_a'+W_bW_b)~.
\eeq
If there is no matter charged under the hidden sector U(1), the hidden sector photon is decoupled and its presence is only felt through a shift  \cite{Holdom:1990xp} in the hypercharge gauge coupling  $\alpha_Y\rightarrow \alpha_Y /(1-\chi)$.
The only physical impact is a perturbation in the precision of gauge coupling unification if $\chi$ is large, which we shall not dwell upon. Whilst the hidden sector photon can be completely decoupled in the absence of light fields charged under the hidden sector U(1), the associated photino will still mix with the visible sector gauginos. This mixing between gaugino states is induced by kinetic mixing and off-diagonal terms in the gaugino mass matrix and appears in the Lagrangian in the form \cite{Arvanitaki:2009hb}
\beq
\mathcal{L}\supset iZ_{ab}\lambda_a^\dagger\slashed\partial\lambda_b+M_{ab}\lambda_a\lambda_b~,
\eeq
where $Z_{ab}$ accounts for the kinetic mixing, $M_{ab}$ is gaugino mass matrix from SUSY breaking and here the indices  $a,b$ run over the bino and the photini states. As previously, the kinetic term can be made canonical by a field redefinition $\lambda_a\rightarrow \lambda_a'=P_{ab}^{-1}\lambda_b$ such that
\beq
\mathcal{L}\supset i\lambda_a'{}^\dagger\slashed\partial\lambda'_a+M'_{ab}\lambda_a'\lambda_b'~,
\eeq
for $M'_{ab}=P_{ac}^{\dagger}M_{cd}P_{db}$. Provided that the original $M_{ab}$ is not proportional to the gauge kinetic mixing matrix, there will be non-zero mixing between the bino and photini states. From this we expect the mixing between a  hidden photino and the neutralino to be parametrically 
\beq
\epsilon\sim 
10^{-7} \times \left( \frac{\chi}{1}\right)
\left(\frac{m_{\widetilde{\gamma}'}}{10~\rm{keV}}\right)
\left(\frac{100~\rm{GeV}}{m_{\widetilde{\chi}}}\right)~.
\label{aa}
\eeq
The scenario in which the hidden photon can be decoupled is of particular interest as it allows for sizeable mixing between the hidden photino and bino, whilst circumventing the constraints coming from the associated kinetic mixing between the visible and hidden sector vector bosons. For a recent analysis of the limits on mixing between gauge bosons see e.g.~\cite{Essig:2013goa}.
 
 In what follows we shall restrict our attention to the simplest scenario with a single hidden sector photino, but in principle there could be many such states with similar or hierarchical masses. Also, we note in passing that the presence of photini could potentially have an impact on collider phenomenology, as studied in \cite{Arvanitaki:2009hb,Baryakhtar:2012rz}.
Supposing that a hidden sector photino is the LSP (as it is a fermion), it is reasonable to suppose that the dominant decay channel for this state is to a neutrino-photon pair via mixing with the neutral gauginos of the MSSM. Hence, in analogy with the previous scenario involving a bino LSP, the lifetime of the hidden photino is 
\beq
\tau_{\widetilde{\gamma}'}\simeq
10^{27}~{\rm s}
&\left(\frac{10^{-7}}{\epsilon}\right)^2
\left(\frac{10^{-2}}{\lambda}\right)^2\\
&\times
\left(\frac{m_{\widetilde{f}}}{500~{\rm GeV}}\right)^4
\left(\frac{7~{\rm keV}}{m_{\widetilde{\gamma}'}}\right)^3.
\label{a}
\eeq
As the decay rate is further suppressed by the mixing parameter $\epsilon$, the coefficient $\lambda$ of the $R$-parity violating operator can be substantially larger than in the bino case whilst ensuring that the LSP is cosmologically stable and with a lifetime suitable to account for the 3.5 keV X-ray line. Similarly, an expression can be obtained for the case that the dominant decay width arises from decays involving a sneutrino VEV, by dressing eq.~(\ref{snu}) with the factor $\epsilon^{-2}$ to account for the $\widetilde{\gamma}'\text{-}\widetilde{B}$ mixing. This yields a hidden photino LSP lifetime of order
\beq 
\tau_{\widetilde{\gamma}'}\simeq
10^{28}~{\rm s}
&\left(\frac{10^{-7}}{\epsilon}\right)^2
\left(\frac{m_{\widetilde{l}}}{100~{\rm GeV}}\right)^4\\
&\times
\left(\frac{1~{\rm MeV}}{\langle\widetilde{\nu}\rangle}\right)^2
\left(\frac{10~{\rm keV}}{m_{\widetilde{\gamma}'}}\right)^3.
\label{as}
\eeq
With weak scale gauginos, the sneutrino VEVs of order $\langle\widetilde{\nu}\rangle\sim1~{\rm MeV}$ lead to sub-eV contributions to the neutrino masses, cf.~eq.~(\ref{nu}).

As discussed previously, obtaining the correct relic density for bino dark matter is non-trivial. The hidden sector photino will also be typically overproduced for $\epsilon\gtrsim10^{-6}$, in which case the photino relic density is given by
\cite{Arvanitaki:2009hb} 
\beq
\Omega_{\widetilde{\gamma}'}h^2\simeq10^{6}\times\left(\frac{m_{\widetilde{\gamma}'}}{10~{\rm keV}}\right)^3\left(\frac{\epsilon}{10^{-3}}\right)^2~.
\eeq
However, if the mixing is small $\epsilon\lesssim10^{-6}$, the hidden sector will never be in thermal contact with the visible sector \cite{Arvanitaki:2009hb}. By inspection of eq.~(\ref{aa}), such small mixing is natural due to the hierarchy between the keV hidden photino, as required to produce X-ray signals, and  weak-scale MSSM gauginos. 
If the inflaton decays preferentially  (similar to \cite{Hall:2009bx}) to the MSSM states, with suppressed couplings to the hidden sector, then substantially fewer photinos will be produced than expected from thermal production.  Neglecting the weak portal interaction the hidden sector is a non-interacting theory and thus the number density of the hidden photino is set primarily by the number produced the inflaton decay. Thus the correct dark matter relic density can be obtained via tuning, somewhat similar to \cite{Chung:1998zb}. Settings in which reheating occurs preferentially have been studied in the context of certain string constructions \cite{Cicoli:2010yj,Cicoli:2010ha}.

A further abundance of photinos will always be generated via thermal freeze-in \cite{Hall:2009bx} due to energy `leaking' from the visible sector to the hidden sector. This can potentially lead to overproduction of the photino and for values of $\epsilon$ which give suitable lifetimes to produce observable signals, if the visible sector reheat temperature is greater than the mass scale of the MSSM superpartners $\widetilde{m}$, then typically the photino will be overproduced. For the case $T_{\rm RH}>\widetilde{m}$ the yield is parametrically
\beq
Y_{\widetilde{\gamma}'}\sim\epsilon^2\left(\frac{M_{\rm Pl}}{\widetilde{m}}\right),
\eeq
and, in which case, the hidden sector photino is overproduced as the yield is related to the relic density as follows
\beq
\Omega_{\widetilde{\gamma}'}h^2\sim 0.1\times
\left(\frac{Y_{\widetilde{\gamma}'}}{10^{-5}} \right)
\left(\frac{m_{\widetilde{\gamma}'}}{10~{\rm keV}} \right)~.
\eeq
On the other hand, if $T_{\rm RH}<\widetilde{m}$, then freeze-in only proceeds via higher dimension operators which connect the hidden sector photinos and the SM states, due to integrating out some heavy superpartners. This leads to a photino yield dependent on  the reheat temperature of the visible sector of the form
\beq
Y_{\widetilde{\gamma}'}
\sim
10^{-5}
\left(\frac{\epsilon}{10^{-7}}\right)^2
\left(\frac{T_{\rm RH}}{1~{\rm GeV}}\right)
\left(\frac{100~{\rm GeV}}{\widetilde{m}}\right)^2~.
\label{Y}
\eeq
In this case, if the correct relic density is set by inflaton decay, the freeze-in abundance can be sufficiently small that it is not then overproduced. Furthermore, in the case that the inflaton does not couple to this sector, for appropriate parameter choices (such as those indicated) the observed dark matter relic density can be achieved directly through this UV freeze-in mechanism. Thus in order for this scenario to work it is crucial that the reheat temperature is below the superpartner mass scale.

Moreover, just as 3-cycles in the internal space are associated to 4d vector bosons, pseudoscalar states --string axions-- arise from integrals of $C_4$ over various 4-cycles \cite{Arvanitaki:2009fg}. It is conceivable that one of these pseudoscalars could play the role of the QCD axion \cite{Shifman:1979if,Kim:1979if}. Assuming this is the case, the indirect indication of a light axion state (as suggested by the experimental requirement that the QCD $\theta$-parameter is near zero) motivates the possibility of a multitude of additional axion states and photini, coined the {\em string axiverse} \cite{Arvanitaki:2009fg}.

\section{Concluding remarks}

We have discussed the prospect for generating keV X-ray signals in SUSY extensions of the Standard Model, with specific reference to the 3.5 keV line recently reported in analysis of the observations by the XMM-Newton X-ray Telescope  \cite{Bulbul:2014sua,Boyarsky:2014jta}. Given that SUSY is the leading candidate for a framework of physics beyond the Standard Model, we believe that it is interesting to consider how such a signal might arise in this setting. We have highlighted the possibility of X-ray signals being generated by the decays of light LSP dark matter via $R$-parity violating operators and argued that such scenarios are viable and motivated, although in some cases require additional model building or some amount of fine-tuning. The gravitino is one of the best motivated light SUSY states, but generically we have argued that it can not lead to signatures of this type. We have proposed rather that scenarios involving keV binos or hidden sector photinos could account for this signal. Interestingly, both these models are sensitive to $T_{\rm RH}$ and typically to be successfully realised it is required that the maximum temperature after reheating is lower than a TeV.

We note in passing that whilst the axino is also a good candidate for a light LSP, similar to the gravitino, a keV axino is typically too long lived to account for the 3.5 keV line. Comparing with the model of e.g.~\cite{Endo:2013si}, the axino lifetime is parametrically
\beq
\tau_{\widetilde{a}}\sim
5\times10^{29}~{\rm s}
&\left(\frac{m_{\widetilde{a}}}{7~{\rm keV}}\right)^{-3}
\left(\frac{f_{a}}{10^8~{\rm GeV}}\right)^2\\
&\times
\left(\frac{m_{\widetilde{B}}}{100~{\rm GeV}}\right)^2
\left(\frac{\langle\widetilde{\nu}\rangle/v}{10^{-6}}\right)^2~.
\eeq
However, as noted in \cite{Kong:2014gea,Choi:2014tva} for extreme parameter choices -- with $f_a$, $m_{\widetilde{B}}$ and $\langle\widetilde{\nu}\rangle$ at the edge of experimental exclusion for most minimal models -- axino interpretations of eq.~(\ref{1}) can be constructed. Aside from these tensions with experimental constraints, this scenario may also be disfavoured from a theoretical stand point  \cite{Cheung:2011mg} as, in the absence of fine tuning or sequestering, the axino mass is expected to be $m_{\widetilde{a}}\gtrsim m_{3/2}$ \cite{Cheung:2011mg}. Thus ensuring that the axino is the LSP requires some model building. Given these considerations we do not discuss this scenario in greater detail.

In closing, we highlight that there is a potential opportunity to distinguish between interpretations involving axions \cite{axion,Jaeckel:2014qea,Lee:2014xua,Cicoli}, and those invoking alternative light dark matter candidates (like the bino or hidden photino) using precision measurements of $\Delta N_{\rm eff}=3\rho_{\rm hidden}/ \rho_{\nu}$.  As typically axions are dominantly produced non-relativistically via the misalignment mechanism   \cite{Shifman:1979if,Kim:1979if}, their contribution to $N_{\rm eff}$ is negligible, in contrast to keV scale thermally produced dark matter. It is projected that upcoming experiments \cite{Abazajian:2013oma} will be sensitive to percent-level changes in $N_{\rm eff}$  and thus should provide some insight regarding the nature of any light hidden sector states.

  Further study of this 3.5 keV X-ray line is certainly warranted. A strong confirmation of this signal, particularly in conjunction with the observation of a factional increase in $N_{\rm eff}$ would be an exciting signal of physics beyond the Standard Model, possibly in the guise of supersymmetry.

{\bf Acknowledgements:}
We are grateful to Antonio Delgado, Zhaofeng Kang, Adam Martin,  Jessie Shelton, and Yue Zhang for useful interactions and the PRD referee for helpful comments. JU is grateful for the hospitality of the Department of Physics at UIUC. This research was supported by the National Science Foundation under Grant No.~PHY-1215979.


\begin{thebibliography}{}   

    

\bibitem{Bulbul:2014sua}
  E.~Bulbul, M.~Markevitch, A.~Foster, R.~K.~Smith, M.~Loewenstein and S.~W.~Randall,
  {\em Detection of An Unidentified Emission Line in the Stacked X-ray spectrum of Galaxy Clusters,} 
Astrophys.\ J.\  {\bf 789} (2014) 13 [1402.2301].
  
\bibitem{Boyarsky:2014jta}
  A.~Boyarsky, O.~Ruchayskiy, D.~Iakubovskyi and J.~Franse,
  {\em An unidentified line in X-ray spectra of the Andromeda galaxy and Perseus galaxy cluster,}
  [1402.4119].
  
\bibitem{Ishida:2014dlp}
  H.~Ishida, K.~S.~Jeong and F.~Takahashi,
  {\em 7 keV sterile neutrino dark matter from split flavor mechanism,}
    Phys.\ Lett.\ B {\bf 732} (2014) 196 [1402.5837].
  
\bibitem{Finkbeiner:2014sja}
  D.~P.~Finkbeiner and N.~Weiner,
  {\em An X-Ray Line from eXciting Dark Matter,}
  [1402.6671].

\bibitem{axion}
  T.~Higaki, K.~S.~Jeong and F.~Takahashi,
  {\em The 7 keV axion dark matter and the X-ray line signal,}
  Phys.\ Lett.\ B {\bf 733} (2014) 25 [1402.6965].
  
  
\bibitem{Jaeckel:2014qea}
  J.~Jaeckel, J.~Redondo and A.~Ringwald,
 {\em A 3.55 keV hint for decaying axion-like particle dark matter,}
  Phys.\ Rev.\ D {\bf 89} (2014) 103511 [1402.7335].
  
\bibitem{Lee:2014xua}
  H.M.~Lee, S.C.~Park and W.-I.~Park,
  {\em Cluster X-ray line at $3.5\,{\rm keV}$ from axion-like dark matter,}
  [1403.0865].

   
\bibitem{Cicoli}
  M.~Cicoli, J.~P.~Conlon, M.~C.~D.~Marsh and M.~Rummel,
  {\em A 3.55 keV Photon Line and its Morphology from a 3.55 keV ALP Line,}
  [1403.2370].
  
\bibitem{Krall:2014dba}
  R.~Krall, M.~Reece and T.~Roxlo,
  {\em Effective field theory and keV lines from dark matter,}
  [1403.1240].
  

\bibitem{Aisati:2014nda}
  C.~\"im.~E.~Aisati, T.~Hambye and T.~Scarna,
  {\em Can a millicharged dark matter particle emit an observable gamma-ray line?,}
  [1403.1280].

  
\bibitem{Frandsen:2014lfa}
  M.~Frandsen, F.~Sannino, I.~M.~Shoemaker and O.~Svendsen,
    {\em X-ray Lines from Dark Matter: The Good, The Bad, and The Unlikely,}
  JCAP {\bf 1405} (2014) 033 [1403.1570].
  
\bibitem{Abazajian:2014gza}
  K.~N.~Abazajian,
  {\em Resonantly-Produced 7 keV Sterile Neutrino Dark Matter Models and the Properties of Milky Way Satellites,}
  Phys.\ Rev.\ Lett.\  {\bf 112} (2014) 161303 [1403.0954].
  

\bibitem{Kong:2014gea}
  K.~Kong, J.~-C.~Park and S.~C.~Park,
  {\em X-ray line signal from 7 keV axino dark matter decay,}
  Phys.\ Lett.\ B {\bf 733} (2014) 217 [1403.1536].
 
\bibitem{Choi:2014tva}
  K.~-Y.~Choi and O.~Seto,
  {\em X-ray line signal from decaying axino warm dark matter,}
  Phys.\ Lett.\ B {\bf 735} (2014) 92 [1403.1782].
  
  
\bibitem{Nakayama:2014ova}
  K.~Nakayama, F.~Takahashi and T.~T.~Yanagida,
  {\em The 3.5 keV X-ray line signal from decaying moduli with low cutoff scale,}
  Phys.\ Lett.\ B {\bf 735} (2014) 338 [1403.1733].
  
\bibitem{Baek:2014qwa}
  S.~Baek and H.~Okada,
  {\em 7 keV Dark Matter as X-ray Line Signal in Radiative Neutrino Model,}
  [1403.1710].
  
\bibitem{Abe:2011ts}
  K.~Abe, 
  T.~Abe, H.~Aihara, Y.~Fukuda, Y.~Hayato, K.~Huang, A.~K.~Ichikawa and M.~Ikeda
   {\it et al.},
  {\em Letter of Intent: The Hyper-Kamiokande Experiment},
  [1109.3262].
  
  
\bibitem{Hikasa:1992je}
  K.~Hikasa {\it et al.}  [Particle Data Group],
  {\em Review of particle properties. Particle Data Group,}
  Phys.\ Rev.\ D {\bf 45} (1992) S1
   [Erratum-ibid.\ D {\bf 46} (1992) 5210].
  
  

\bibitem{Jungman:1995df}
  G.~Jungman, M.~Kamionkowski and K.~Griest,
  {\em Supersymmetric dark matter,}
  Phys.\ Rept.\  {\bf 267} (1996) 195
  [hep-ph/9506380].
  

\bibitem{Hall:1983id}
  L.~J.~Hall and M.~Suzuki,
  {\em Explicit R-Parity Breaking in Supersymmetric Models,}
  Nucl.\ Phys.\ B {\bf 231} (1984) 419.
  
  

\bibitem{Dawson:1985vr}
  S.~Dawson,
  {\em R-Parity Breaking in Supersymmetric Theories,}
  Nucl.\ Phys.\ B {\bf 261} (1985) 297.

 
  \bibitem{Dreiner:1997uz}
  H.~K.~Dreiner,
 {\em An Introduction to explicit R-parity violation,}
  In *Kane, G.L. (ed.): Perspectives on supersymmetry II* 565-583
  [hep-ph/9707435].


\bibitem{Barbier:2004ez}
  R.~Barbier, C.~Berat, M.~Besancon, M.~Chemtob, A.~Deandrea, E.~Dudas, P.~Fayet and S.~Lavignac {\it et al.},
  {\em R-parity violating supersymmetry,}
  Phys.\ Rept.\  {\bf 420} (2005) 1
  [hep-ph/0406039].
  

\bibitem{Viel:2005qj}
  M.~Viel, J.~Lesgourgues, M.~G.~Haehnelt, S.~Matarrese and A.~Riotto,
  {\em Constraining warm dark matter candidates including sterile neutrinos and light gravitinos with WMAP and the Lyman-alpha forest,}
  Phys.\ Rev.\ D {\bf 71} (2005) 063534
  [astro-ph/0501562].
   
   
\bibitem{Takayama:2000uz}
  F.~Takayama and M.~Yamaguchi,
  {\em Gravitino dark matter without R-parity,}
  Phys.\ Lett.\ B {\bf 485} (2000) 388
  [hep-ph/0005214].
 
\bibitem{Ishiwata:2008cu}
  K.~Ishiwata, S.~Matsumoto and T.~Moroi,
  {\em High Energy Cosmic Rays from the Decay of Gravitino Dark Matter,}
  Phys.\ Rev.\ D {\bf 78} (2008) 063505,
  [0805.1133].
  
  
  

\bibitem{Buchmuller:2007ui}
  W.~Buchmuller, L.~Covi, K.~Hamaguchi, A.~Ibarra and T.~Yanagida,
  {\em Gravitino Dark Matter in R-Parity Breaking Vacua,}
  JHEP {\bf 0703} (2007) 037,
  hep-ph/0702184.


\bibitem{Endo:2009by}
  M.~Endo and T.~Shindou,
  {\em R-parity Violating Right-Handed Neutrino in Gravitino Dark Matter Scenario,}
  JHEP {\bf 0909} (2009) 037
  [0903.1813].
  

    
\bibitem{Bobrovskyi:2010ps}
  S.~Bobrovskyi, W.~Buchmuller, J.~Hajer and J.~Schmidt,
  {\em Broken R-Parity in the Sky and at the LHC,}
  JHEP {\bf 1010} (2010) 061
  [1007.5007].


\bibitem{Bajc:2010qj}
  B.~Bajc, T.~Enkhbat, D.~K.~Ghosh, G.~Senjanovic and Y.~Zhang,
  {\em MSSM in view of PAMELA and Fermi-LAT,}
  JHEP {\bf 1005} (2010) 048
  [1002.3631].

  
\bibitem{Bomark:2014yja}
  N.~-E.~Bomark and L.~Roszkowski,
  {\em The 3.5 keV X-ray line from decaying gravitino dark matter,}
  Phys.\ Rev.\ D {\bf 90} (2014) 011701 [1403.6503].
  

\bibitem{Lola:2007rw}
  S.~Lola, P.~Osland and A.~R.~Raklev,
  {\em Radiative gravitino decays from R-parity violation,}
  Phys.\ Lett.\ B {\bf 656} (2007) 83
  [0707.2510].
  
    

\bibitem{Profumo:2008yg}
  S.~Profumo,
 {\em Hunting the lightest lightest neutralinos,}
  Phys.\ Rev.\ D {\bf 78} (2008) 023507,
  [0806.2150].
  
\bibitem{Gogoladze:2002xp}
  I.~Gogoladze, J.~D.~Lykken, C.~Macesanu and S.~Nandi,
  {\em Implications of a massless neutralino for neutrino physics,}
  Phys.\ Rev.\ D {\bf 68} (2003) 073004
  [hep-ph/0211391].


\bibitem{Dreiner:2009ic}
  H.~K.~Dreiner, S.~Heinemeyer, O.~Kittel, U.~Langenfeld, A.~M.~Weber and G.~Weiglein,
  {\em Mass Bounds on a Very Light Neutralino,}
  Eur.\ Phys.\ J.\ C {\bf 62} (2009) 547
  [0901.3485].


\bibitem{Davies:2011mp}
  R.~Davies, J.~March-Russell and M.~McCullough,
 {\em A Supersymmetric One Higgs Doublet Model,}
  JHEP {\bf 1104} (2011) 108
  [1103.1647].
  
    
\bibitem{Fox:2002bu}
  P.~J.~Fox, A.~E.~Nelson and N.~Weiner,
  {\em Dirac gaugino masses and supersoft supersymmetry breaking,}
  JHEP {\bf 0208} (2002) 035
  [hep-ph/0206096].
  
\bibitem{Ellwanger:2009dp}
  U.~Ellwanger, C.~Hugonie and A.~M.~Teixeira,
 {\em The Next-to-Minimal Supersymmetric Standard Model,}
  Phys.\ Rept.\  {\bf 496} (2010) 1
  [0910.1785].
    
\bibitem{Randall:1998uk}
  L.~Randall and R.~Sundrum,
  {\em Out of this world supersymmetry breaking,}
  Nucl.\ Phys.\ B {\bf 557} (1999) 79
  [hep-th/9810155].
  
\bibitem{Giudice:1998xp}
  G.~F.~Giudice, M.~A.~Luty, H.~Murayama and R.~Rattazzi,
  {\em Gaugino mass without singlets,}
  JHEP {\bf 9812} (1998) 027
  [hep-ph/9810442].


\bibitem{de Gouvea:1997tn}
  A.~de Gouvea, T.~Moroi and H.~Murayama,
  {\em Cosmology of supersymmetric models with low-energy gauge mediation,}
  Phys.\ Rev.\ D {\bf 56} (1997) 1281
  [hep-ph/9701244].
          
  
\bibitem{Allanach:1999ic}
  B.~C.~Allanach, A.~Dedes and H.~K.~Dreiner,
  {\em Bounds on R-parity violating couplings at the weak scale and at the GUT scale,}
  Phys.\ Rev.\ D {\bf 60} (1999) 075014
  [hep-ph/9906209].

\bibitem{Barger:2008wn} 
  V.~Barger, P.~Fileviez Perez and S.~Spinner,
 {\em Minimal gauged U(1)(B-L) model with spontaneous R-parity violation,}
  Phys.\ Rev.\ Lett.\  {\bf 102}, 181802 (2009)
  [0812.3661].

  

\bibitem{Takayama:1999pc}
  F.~Takayama and M.~Yamaguchi,
  {\em Pattern of neutrino oscillations in supersymmetry with bilinear R-parity violation,}
  Phys.\ Lett.\ B {\bf 476} (2000) 116
  [hep-ph/9910320].
  
\bibitem{Giudice:2000ex}
  G.~F.~Giudice, E.~W.~Kolb and A.~Riotto,
 {\em Largest temperature of the radiation era and its cosmological implications,}
  Phys.\ Rev.\ D {\bf 64} (2001) 023508
  [hep-ph/0005123].
  
\bibitem{Jedamzik:2006xz}
  K.~Jedamzik,
  {\em Big bang nucleosynthesis constraints on hadronically and electromagnetically decaying relic neutral particles,}
  Phys.\ Rev.\ D {\bf 74} (2006) 103509
  [hep-ph/0604251].
  
\bibitem{D'Eramo:2011ec}
  F.~D'Eramo, L.~Fei and J.~Thaler,
  {\em Dark Matter Assimilation into the Baryon Asymmetry,}
  JCAP {\bf 1203} (2012) 010
  [1111.5615].
    
    
    
\bibitem{Kamionkowski:1999qc}
  M.~Kamionkowski and A.~Kosowsky,
    {\em The Cosmic microwave background and particle physics,}
  Ann.\ Rev.\ Nucl.\ Part.\ Sci.\  {\bf 49} (1999) 77
  [astro-ph/9904108].
  

\bibitem{Boehm:2013jpa}
  C.~Boehm, M.~J.~Dolan and C.~McCabe,
  {\em A Lower Bound on the Mass of Cold Thermal Dark Matter from Planck,}
  JCAP {\bf 1308} (2013) 041
  [1303.6270].

    
\bibitem{Arvanitaki:2009hb}
  A.~Arvanitaki, N.~Craig, S.~Dimopoulos, S.~Dubovsky and J.~March-Russell,
  {\em String Photini at the LHC,}
  Phys.\ Rev.\ D {\bf 81} (2010) 075018
  [0909.5440].
  
\bibitem{Ibarra:2008kn}
  A.~Ibarra, A.~Ringwald and C.~Weniger,
  {\em Hidden gauginos of an unbroken U(1): Cosmological constraints and phenomenological prospects,}
  JCAP {\bf 0901} (2009) 003
  [0809.3196].


    
\bibitem{Holdom:1985ag}
  B.~Holdom,
  {\em Two U(1)'s and Epsilon Charge Shifts,}
  Phys.\ Lett.\ B {\bf 166} (1986) 196.
  
\bibitem{Dienes:1996zr}
  K.~R.~Dienes, C.~F.~Kolda and J.~March-Russell,
  {\em Kinetic mixing and the supersymmetric gauge hierarchy,}
  Nucl.\ Phys.\ B {\bf 492} (1997) 104
  [hep-ph/9610479].
  
      
\bibitem{Holdom:1990xp}
  B.~Holdom,
  {\em Oblique electroweak corrections and an extra gauge boson,}
  Phys.\ Lett.\ B {\bf 259} (1991) 329.


\bibitem{Essig:2013goa}
  R.~Essig, E.~Kuflik, S.~D.~Mcdermott, T.~Volansky and K.~M.~Zurek,
  {\em Constraining Light Dark Matter with Diffuse X-Ray and Gamma-Ray Observations,}
  JHEP {\bf 1311} (2013) 193,
  [1309.4091].
  
  


\bibitem{Baryakhtar:2012rz}
  M.~Baryakhtar, N.~Craig and K.~Van Tilburg,
  {\em Supersymmetry in the Shadow of Photini,}
  JHEP {\bf 1207} (2012) 164
  [1206.0751].

  
\bibitem{Hall:2009bx}
  L.~J.~Hall, K.~Jedamzik, J.~March-Russell and S.~M.~West,
  {\em Freeze-In Production of FIMP Dark Matter,}
  JHEP {\bf 1003} (2010) 080
  [0911.1120].
  
\bibitem{Chung:1998zb}
  D.~J.~H.~Chung, E.~W.~Kolb and A.~Riotto,
  {\em Superheavy dark matter,}
  Phys.\ Rev.\ D {\bf 59} (1998) 023501
  [hep-ph/9802238].
 
\bibitem{Cicoli:2010ha}
  M.~Cicoli and A.~Mazumdar,
  {\em Reheating for Closed String Inflation,}
  JCAP {\bf 1009} (2010) 025,
  [1005.5076].
  
\bibitem{Cicoli:2010yj}
  M.~Cicoli and A.~Mazumdar,
  {\em Inflation in string theory: A Graceful exit to the real world,}
  Phys.\ Rev.\ D {\bf 83} (2011) 063527,
  [1010.0941].


\bibitem{Arvanitaki:2009fg}
  A.~Arvanitaki, S.~Dimopoulos, S.~Dubovsky, N.~Kaloper and J.~March-Russell,
  {\em String Axiverse,}
  Phys.\ Rev.\ D {\bf 81} (2010) 123530
  [0905.4720].
  
  
\bibitem{Kim:1979if}
  J.~E.~Kim,
  {\em Weak Interaction Singlet and Strong CP Invariance,}
  Phys.\ Rev.\ Lett.\  {\bf 43} (1979) 103.
  
\bibitem{Shifman:1979if}
  M.~A.~Shifman, A.~I.~Vainshtein and V.~I.~Zakharov,
  {\em Can Confinement Ensure Natural CP Invariance of Strong Interactions?,}
  Nucl.\ Phys.\ B {\bf 166} (1980) 493.
  
  
  

\bibitem{Endo:2013si}
  M.~Endo, K.~Hamaguchi, S.~P.~Liew, K.~Mukaida and K.~Nakayama,
  {\em Axino dark matter with R-parity violation and 130 GeV gamma-ray line,}
  Phys.\ Lett.\ B {\bf 721} (2013) 111
  [1301.7536].


\bibitem{Cheung:2011mg}
  C.~Cheung, G.~Elor and L.~J.~Hall,
  {\em The Cosmological Axino Problem,}
  Phys.\ Rev.\ D {\bf 85} (2012) 015008
  [1104.0692].


\bibitem{Abazajian:2013oma}
  K.~N.~Abazajian, 
  K.~Arnold, J.~Austermann, B.~A.~Benson, C.~Bischoff, J.~Bock, J.~R.~Bond and J.~Borrill 
  {\it et al.},
  {\em Neutrino Physics from the Cosmic Microwave Background and Large Scale Structure,}
  [1309.5383].
    
\end{thebibliography}
\end{document}